\documentclass[twocolumn,english,showpacs,amssymb,amsmath,prl]{revtex4}
\usepackage[T1]{fontenc}
\usepackage[latin9]{inputenc}
\usepackage{graphicx}

\usepackage{dcolumn}
\usepackage{bm}

\begin{document}

\title{Electrostatic Control of the Evolution from Superconductor to Insulator
in Ultrathin Films of Yttrium Barium Copper Oxide}

\author{Xiang Leng}

\author{Javier Garcia-Barriocanal}

\author{Yeonbae Lee}

\author{A. M. Goldman}

\affiliation{School of Physics and Astronomy, University of Minnesota, Minneapolis,
Minnesota 55455, USA}

\date{\today}
\begin{abstract}
The electrical transport properties of ultrathin
YBa$_{2}$Cu$_{3}$O$_{7-x}$ films have been modified using an
electric double layer transistor configuration employing an ionic
liquid. The films were grown on SrTiO$_{3}$ substrates using high
pressure oxygen sputtering. A clear evolution from superconductor
to insulator was observed in nominally 7 unit cell thick films.
Using a finite size scaling analysis, curves of resistance versus
temperature, $R\left(T\right),$ over the temperature range from 6K
to 22K were found to collapse onto a single scaling function,
which suggests the the presence of a quantum critical point.
However the scaling failed at the lowest temperatures suggesting
the presence of an additional phase between the superconducting
and insulating regimes.
\end{abstract}


\maketitle
Superconductor-insulator (SI) transitions at zero temperature induced
by varying external parameters such as thickness, magnetic field and
carrier concentration are examples of quantum phase transitions (QPTs)\cite{SondhiRMP,GantmakherSIQPT}.
A signature of a QPT at nonzero temperature is the success of finite
size scaling in describing the data\cite{GoldmanPhysicsToday,AliPRLMoGeMagneticfield,KevinPRL2005}.
Basically, the resistance of a 2D system near a quantum critical point
collapses onto a single scaling function $R=R_{c}f(\delta T^{-\nu z})$,
where $R_{c}$ is the critical resistance, $\delta$ is tuning parameter,
$T$ is the temperature, $\nu$ is the correlation length critical
exponent and z is the dynamic critical exponent\cite{ChaFisherPRB1991,FisherPRB1989,FisherPRL1990disorder,FisherPRL1990Rc}.

For high-$T_{c}$ cuprates, such as YBa$_{2}$Cu$_{3}$O$_{7-x}$
(YBCO), since thickness cannot be varied
continuously\cite{WangYBCOthickness,SalluzzoNBCO,OrgianiPRLhighTcRc},
and the upper critical field $H_{c2}$ is
huge\cite{AndoLSCOHighfield,SteinerHighTcSIhighField,SeidlerSIinYBCO}
tuning carrier concentration would be the most practical approach.
There are two ways to do this, by chemical doping or by
electrostatically charging. The disadvantage of the former is that
it can introduce structural or chemical changes and cannot be
tuned continuously\cite{NicolasPRLYBCOSI1}. On the other hand,
electrostatic charging is promising since it keeps the structure
fixed and it's continuous and
reversible\cite{AhnScience1999GBCO,MannhartSSTreview,TandaSIinNCCO,AhnRMP}.
Moreover, a large effect is expected in high-$T_{c}$ cuprates due
to their much lower carrier densities compared with those of
conventional metals. However, most electrostatic charging
experiments with cuprates show only small changes in
$T_{c}$\cite{RufenachtSupercluidDensityinLSCO,MattheyNBCOonSTO,SalluzzoSST,SalluzzoPRB2008},
suggesting that the situation is more complicated. Work using
SrTiO$_3$ (STO) as the dielectric indicates that although the
carrier concentration is changed by the order of $10^{13}{\rm
{cm^{-2}}}$, this change is not large enough to traverse from
insulator to superconductor, as the induced carriers may reside on
the CuO$_{x}$ chains while the CuO$_{2}$ planes are only
indirectly
affected\cite{SalluzzoPRLNBCOCuOxchain,SalluzzoPRB2007SI}.

\begin{figure}[t]
 \includegraphics[width=0.36\textwidth,height=0.3\textwidth]{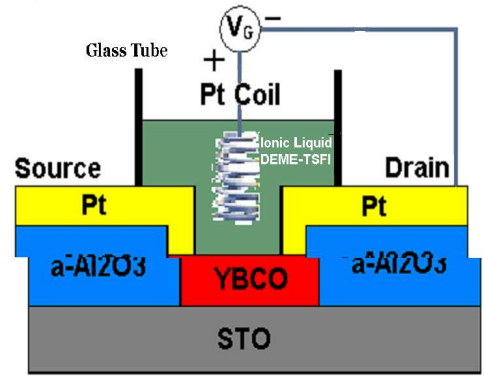}
\caption{(color online) Cartoon of the FET structure (side view). A patterned
amorphous $Al_{2}O_{3}$ film was deposited to define the geometry
before YBCO film growth.This enabled us to avoid post-deposition patterning.
Pt electrodes were deposited through a shadow mask. The ionic liquid
was filled into a small glass tube glued on top of the structure.
\label{sketch}}

\end{figure}

Recently, ionic liquids have been used to make FET-like devices to
induce large carrier concentration changes in inorganic materials
such as ZrNCl, STO, ZnO and
YBCO\cite{YeNature,YeonbaeSTO,ShimotaniZnO,DhootYBCO}. The devices
are electric double layer transistors (EDLTs), since in response
to the electric field, ions are driven to sample surfaces to form
an interfacial double layer with the induced carriers. The gap
between these two layers is of the order of nanometers. The
capacitance is huge and the induced sheet carrier density can be
as high as $8\times10^{14}cm^{2}$. With such high carrier
concentration, superconductivity was induced in ZrNCl and STO
samples and a T$_{c}$ shift of about 50K was observed in YBCO thin
films. However, no continuous transition from insulator to
superconductor has thus far been found in YBCO. The reason is that
films as thin as 10 unit cells (UC) are still too thick. The
Thomas-Fermi screening length in YBCO is short, typically several
Angstroms. Thus the order of only one unit cell (UC) will be
affected by gating. In a thick sample, the proximity effect will
dominate and no transition will be seen. Recently, Bollinger
\textit{et al.} reported superconductor-insulator transition in a
214 compound La$_{2-x}$Sr$_{x}$CuO$_{4}$ using an EDLT
configuration\cite{BollingerLSCO}. Here we report a clear
electrostatically induced transition between superconducting and
insulating behavior in YBCO, which is a 123 compound. The 214 and
123 compounds have different structures.

Films were grown on (001) oriented SrTiO$_{3}$ substrates using a
high pressure oxygen sputtering system. Previous work has shown that
this technique provides high quality epitaxial ultrathin YBCO films
\cite{JacoboYBCOPRL1999,JacoboYBCOPRL2001}. A sketch of the device
is shown in Fig.~\ref{sketch}. Amorphous Al$_{2}$O$_{3}$ films
were deposited as masks, and the substrates were etched in buffered
(10:1) HF solution and annealed in an O$_{2}$/O$_{3}$ mixture for
6 hours at 750 $\mathrm{^{o}}\mathrm{C}$. Atomic force microscopy
(AFM) was used to make sure that the substrate surfaces were clean
and were TiO$_{2}$ terminated. The oxygen pressure during deposition
was 2.0 mbar and the substrate temperature was 900 $\mathrm{^{o}}\mathrm{C}$.
After deposition, films were cooled in an 800 mbar O$_{2}$ atmosphere
and annealed at 500$\mathrm{^{o}}\mathrm{C}$ for 30 minutes. Films
were characterized \textit{ex-situ} by AFM and X-ray diffraction.
Thickness was measured using X-ray reflectivity. A series of samples
with thicknesses ranging from 5 to 10 UC were fabricated and measured
using standard four-probe techniques. Films thinner than or equal
to 6 UC were insulating whereas thicker ones were superconducting.
This suggests that the first 5 to 6 UC are insulating and that the
7 UC thick film has a superconducting layer that is actually only
1 to 2 UC thick.

Ultrathin YBCO films are sensitive and react with most chemicals so
we didn't carry out any processing after film growth. Films were covered
with ionic liquid (IL) and were quickly cooled down to 240K. The IL
used was N,N-diethly-N-(2-methoxyethyl)-N-methylammonium bis(trifluoromethylsulphonyl)-imide,
(DEME-TFSI). This condenses into a rubber-like state at 240K, a temperature
at which most chemical reactions are suppressed. If the device is
kept at room temperature for several hours with ionic liquid on it,
we see an increase in resistance and a drop in $T_{c}$, possibly
due to chemical reaction. However, after cooling to 240K and being
kept below this temperature, no changes are seen over several days.
For measurements, the gate voltage was changed at 240K where it is
held for 1 hour before cooling down. The gate voltage was kept constant
during measurements.

\begin{figure}[t]
 \includegraphics[width=0.36\textwidth,height=0.3\textwidth]{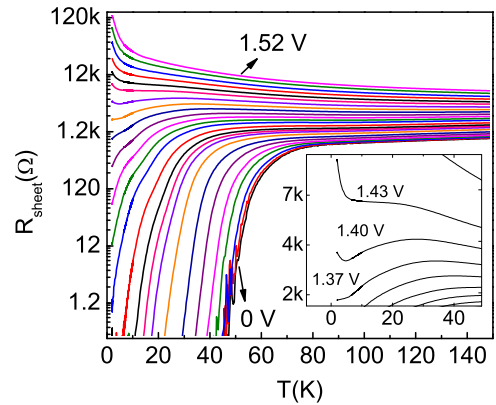}
\caption{(color online) Logarithm of the sheet resistance vs temperature with
gate voltage varying from 0 to 1.52V. From bottom to top: 0, 0.3,
0.5, 0.7, 0.8, 0.9, 1, 1.1, 1.12, 1.14, 1.17, 1.20, 1.25, 1.27, 1.30,
1.33, 1.35, 1.37, 1.40, 1.43, 1.44, 1.46, 1.48, 1.50, 1.52 (V). Inset:
enlarged low temperature part near the transition between the superconducting
and insulating regimes. \label{RT}}

\end{figure}

Resistance vs temperature, $R\left(T\right)$, curves at various gate
voltages of a 7 UC thick YBCO sample are shown in Fig.~\ref{RT}.
The sample starts as a superconductor with $T_{c}^{onset}\sim77K$.
Here $T_{c}^{onset}$ is taken to be the temperature at which the
sheet resistance falls to 90\% of its normal value. We realize an
insulator with a gate voltage ($V_{G}$) of only 1.52V. We notice
that $T_{c}^{onset}$ is a nonlinear function of $V_{G}$. There is
a V$_{G}$ threshold of about 0.3V below which almost no change can
be seen. Then, $T_{c}^{onset}$ drops by about 30K as V$_{G}$ increases
from 0.5V to 1.1V, which is about 5K/0.1V. However, as V$_{G}$ increases
from 1.1V to 1.25V, a change of 0.2V to 0.3V can induce a $T_{c}^{onset}$
shift of about 5K. This nonlinearity suggests that the gate voltage
cannot be used as a tuning parameter for quantitative analysis.

As $V_{G}$ increases to 1.37V, $R\left(T\right)$ curves flatten
out at the lowest temperature then undergo a small upturn at $V_{G}=$1.40V
although they initially decrease as temperature decreases. Finally
$R\left(T\right)$ evolves to an insulating state as $V_{G}$ increases
further, as can be seen clearly in the inset of Fig.~\ref{RT}. This
low temperature behavior is similar to that observed in amorphous
MoGe films and granular superconductors\cite{AliPRLMoGeMagneticfield,Allen2DGranular,SpivakTheoryPRB,ChenRTupturnPRB2009,DasDoniachPRBBosemetal}
and will dramatically affect any quantitative scaling analysis. We
suggest that this is evidence of a mixed phase separating the superconducting
and insulating regimes.

\begin{figure}[t]
 \includegraphics[width=0.36\textwidth,height=0.3\textwidth]{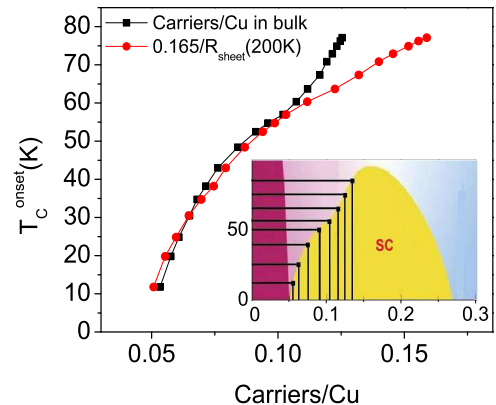}
\caption{(color online) Carrier concentration calculated using 0.165/R$_{sheet}$(200K)
(red dots) compared to that derived from the phase diagram of bulk
high-T$_{c}$ cuprates (black squares). The inset shows how the carrier
concentration was determined from the bulk phase diagram using measured
values of $T_{c}^{onset}$ at different gate voltages. \label{carrier}}

\end{figure}

\begin{figure}[t]
 \includegraphics[width=0.36\textwidth,height=0.3\textwidth]{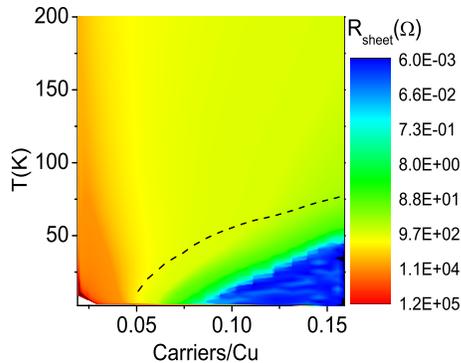}
\caption{(color online) Color plot of the resistance vs temperature and calculated
carrier concentration. This can be interpreted as a phase diagram.
Different colors represent different sheet resistances. The dash line
is $T_{c}^{onset}$. The white area in the bottom left hand corner
is a regime of with nonlinear I-V curves. \label{phase}}

\end{figure}

For further quantitative analysis of the data, we need the induced
carrier concentration. As discussed above, gate voltage is not simply
related to the carrier concentration and the Hall resistance can't
be used because of the complicated electronic properties of YBCO.
We assume that Drude behavior is found at high temperatures and take
1/R$_{sheet}$(T=200K) to be proportional to the carrier concentration
measured as carriers per in-plane Cu atom. Comparing this with the
carriers per in-plane Cu atom derived from the phase diagram of bulk
YBCO\cite{YBCOphase}, we find that they match with a constant coefficient
0.165 as plotted in Fig.~\ref{carrier}. This is especially true
in the transition regime which is the focus of quantitative analysis.

Using measured $R\left(T\right)$ curves and calculated carrier concentrations,
a phase diagram can be constructed and is plotted as Fig.~\ref{phase}.
It is similar to the phase diagram of bulk high-$T_{c}$ cuprates
derived from chemical doping. Here it is measured continuously and
perhaps more accurately, especially in the transition and insulating
regimes. The data plotted here are in the linear regime of the I-V
curves. There are nonlinear I-V curves deep in the insulating regime
which are denoted by white in the figure.

\begin{figure}[t]
 \includegraphics[width=0.36\textwidth,height=0.3\textwidth]{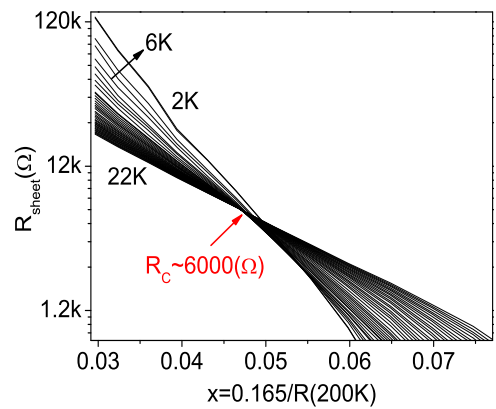}
\caption{Isotherms of $R\left(x\right)$ at temperatures ranging from 2K to
22K.. The carrier concentration $x$ is calculated using the phenomenological
relation x=0.165/R$_{sheet}(200K)$. The arrow indicates the crossing
point where R$_{c}$=6.0k$\Omega$ and x$_{c}$=0.048 carriers/Cu.\label{Rc}}

\end{figure}

When a quantum critical point is approached, resistance isotherms
as a function of the carrier concentration $x$ should cross at the
critical resistance $R_{c}$ and all resistance data should collapse
onto a single scaling function. Figure~\ref{Rc} shows isotherms
from 2K to 22K. A clean crossing point can be seen if we neglect the
lowest temperature isotherms. The critical resistance per square is
about 6.0k$\Omega$, which is very close to the quantum resistance
given by $R_{Q}=h/(2e)^{2}=6.45k\Omega$ considering that imprecise
shadow masks were used to define the four-terminal configuration.
The critical carrier concentration is x$_{c}$=0.048 carriers/Cu,
close to that derived from the general phase diagram (0.05). But this
is not surprising given that the bulk phase diagram was used to calibrate
the carrier concentration. %

\begin{figure}[t]
 \includegraphics[width=0.36\textwidth,height=0.3\textwidth]{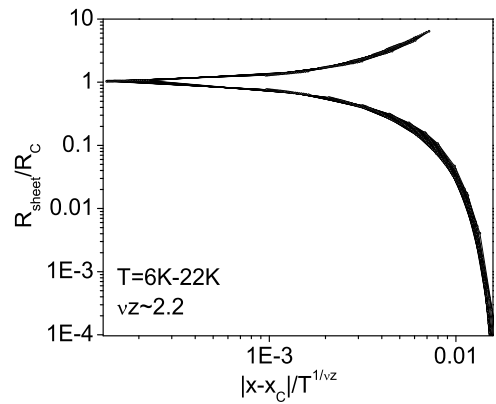}
\caption{Finite size scaling analysis of $R\left(T\right)$ using the calculated
carrier concentration x=0.165/R$_{sheet}(T=200K)$ as tuning parameter.
Only data in the range from 6K to 22K are shown here and the best
collapse is found when $\nu z=2.2$. \label{scale}}

\end{figure}

The results of the finite size scaling analysis are shown in Fig.~\ref{scale}.
All the sheet resistance data from 6K to 22K collapse onto a single
function $R_{sheet}=R_{c}f(|x-x_{c}|T^{-\nu z})$ if $\nu z=2.2$.
Assuming z=1 \cite{Herbutquantumcriticalpoint}, the value $\nu=2.2$
is close to the universality class of metal-insulator transition in
anisotropic 2D system (7/3)\cite{MarcMItransition} or quantum percolation
model (2.43)\cite{DunghaiLeeQuantumPercolation}. Scaling breaks down
for data below 6K, which is not surprising due to the nonmonotonic
behavior of $R\left(T\right)$ at low temperature at gate voltages
in the transition regime.

In summary, a transition from a superconductor to an insulator has
been induced in ultrathin superconducting YBCO films using an EDLT
device configuration. A striking feature of the data is the similarity
between the phase diagram (Fig.~\ref{phase}) and bulk phase diagram.
This is surprising considering the fact that the active layer in our
sample is only one to two UCs and that there are high electric fields
in the double layer. All resistance data collapse onto a single scaling
function with critical exponent $\nu z=2.2$ except the lowest temperature
part. This suggests an approach to a quantum critical point. However,
new physics develops at the lowest temperatures, resulting in curves
similar to those found in granular superconductors. The resultant
SI transition is not direct.

We would like to thank Boris Shklovskii and Yen-Hsiang Lin for
useful discussions. We also would like to acknowledge Shameek Bose
and Christopher Leighton for their assistance with sample
preparation and Chad Geppert for his help with the measurements.
Jacobo Santamaria contributed to the early stages of the work.
Finally we would like to thank Ivan Bozovic for making his results
available in advance of publication. This work was supported by
the National Science Foundation under grant NSF/DMR-0709584. Part
of this work was carried out at the University of Minnesota
Characterization Facility, a member of the NSF-funded Materials
Research Facilities Network via the MRSEC program, and the
Nanofabrication Center which receive partial support from the NSF
through the NNIN program. JGB thanks the Spanish Ministry of
Education for the financial support through the National Program
of Mobility of Human Resources (2008-2011).


\begin{thebibliography}{41}
\expandafter\ifx\csname
natexlab\endcsname\relax\def\natexlab#1{#1}\fi
\expandafter\ifx\csname bibnamefont\endcsname\relax
  \def\bibnamefont#1{#1}\fi
\expandafter\ifx\csname bibfnamefont\endcsname\relax
  \def\bibfnamefont#1{#1}\fi
\expandafter\ifx\csname citenamefont\endcsname\relax
  \def\citenamefont#1{#1}\fi
\expandafter\ifx\csname url\endcsname\relax
  \def\url#1{\texttt{#1}}\fi
\expandafter\ifx\csname
urlprefix\endcsname\relax\def\urlprefix{URL }\fi
\providecommand{\bibinfo}[2]{#2}
\providecommand{\eprint}[2][]{\url{#2}}

\bibitem[{\citenamefont{Sondhi et~al.}(1997)\citenamefont{Sondhi, Girvin,
  Carini, and Shahar}}]{SondhiRMP}
\bibinfo{author}{\bibfnamefont{S.~L.} \bibnamefont{Sondhi}},
  \bibinfo{author}{\bibfnamefont{S.~M.} \bibnamefont{Girvin}},
  \bibinfo{author}{\bibfnamefont{J.~P.} \bibnamefont{Carini}},
  \bibnamefont{and} \bibinfo{author}{\bibfnamefont{D.}~\bibnamefont{Shahar}},
  \bibinfo{journal}{Rev. Mod. Phys.} \textbf{\bibinfo{volume}{69}},
  \bibinfo{pages}{315} (\bibinfo{year}{1997}).

\bibitem[{\citenamefont{Gantmakher and Dolgopolov}(2010)}]{GantmakherSIQPT}
\bibinfo{author}{\bibfnamefont{V.~F.} \bibnamefont{Gantmakher}}
  \bibnamefont{and} \bibinfo{author}{\bibfnamefont{V.~T.}
  \bibnamefont{Dolgopolov}}, \bibinfo{journal}{Physics-Uspekhi}
  \textbf{\bibinfo{volume}{53}}, \bibinfo{pages}{1} (\bibinfo{year}{2010}).

\bibitem[{\citenamefont{Goldman and Markovic}(1998)}]{GoldmanPhysicsToday}
\bibinfo{author}{\bibfnamefont{A.~M.} \bibnamefont{Goldman}} \bibnamefont{and}
  \bibinfo{author}{\bibfnamefont{N.}~\bibnamefont{Markovic}},
  \bibinfo{journal}{Physics Today} \textbf{\bibinfo{volume}{51}},
  \bibinfo{pages}{39} (\bibinfo{year}{1998}).

\bibitem[{\citenamefont{Yazdani and
  Kapitulnik}(1995)}]{AliPRLMoGeMagneticfield}
\bibinfo{author}{\bibfnamefont{A.}~\bibnamefont{Yazdani}} \bibnamefont{and}
  \bibinfo{author}{\bibfnamefont{A.}~\bibnamefont{Kapitulnik}},
  \bibinfo{journal}{Phys. Rev. Lett.} \textbf{\bibinfo{volume}{74}},
  \bibinfo{pages}{3037} (\bibinfo{year}{1995}).

\bibitem[{\citenamefont{Parendo et~al.}(2005)}]{KevinPRL2005}
\bibinfo{author}{\bibfnamefont{K.~A.} \bibnamefont{Parendo}}
  \bibnamefont{et~al.}, \bibinfo{journal}{Phys. Rev. Lett.}
  \textbf{\bibinfo{volume}{94}}, \bibinfo{pages}{197004}
  (\bibinfo{year}{2005}).

\bibitem[{\citenamefont{Cha et~al.}(1991)}]{ChaFisherPRB1991}
\bibinfo{author}{\bibfnamefont{M.-C.} \bibnamefont{Cha}} \bibnamefont{et~al.},
  \bibinfo{journal}{Phys. Rev. B} \textbf{\bibinfo{volume}{44}},
  \bibinfo{pages}{6883} (\bibinfo{year}{1991}).

\bibitem[{\citenamefont{Fisher et~al.}(1989)\citenamefont{Fisher, Weichman,
  Grinstein, and Fisher}}]{FisherPRB1989}
\bibinfo{author}{\bibfnamefont{M.~P.~A.} \bibnamefont{Fisher}},
  \bibinfo{author}{\bibfnamefont{P.~B.} \bibnamefont{Weichman}},
  \bibinfo{author}{\bibfnamefont{G.}~\bibnamefont{Grinstein}},
  \bibnamefont{and} \bibinfo{author}{\bibfnamefont{D.~S.}
  \bibnamefont{Fisher}}, \bibinfo{journal}{Phys. Rev. B}
  \textbf{\bibinfo{volume}{40}}, \bibinfo{pages}{546} (\bibinfo{year}{1989}).

\bibitem[{\citenamefont{Fisher}(1990)}]{FisherPRL1990disorder}
\bibinfo{author}{\bibfnamefont{M.~P.~A.} \bibnamefont{Fisher}},
  \bibinfo{journal}{Phys. Rev. Lett.} \textbf{\bibinfo{volume}{65}},
  \bibinfo{pages}{923} (\bibinfo{year}{1990}).

\bibitem[{\citenamefont{Fisher et~al.}(1990)\citenamefont{Fisher, Grinstein,
  and Girvin}}]{FisherPRL1990Rc}
\bibinfo{author}{\bibfnamefont{M.~P.~A.} \bibnamefont{Fisher}},
  \bibinfo{author}{\bibfnamefont{G.}~\bibnamefont{Grinstein}},
  \bibnamefont{and} \bibinfo{author}{\bibfnamefont{S.~M.}
  \bibnamefont{Girvin}}, \bibinfo{journal}{Phys. Rev. Lett.}
  \textbf{\bibinfo{volume}{64}}, \bibinfo{pages}{587} (\bibinfo{year}{1990}).

\bibitem[{\citenamefont{Wang et~al.}(1991)}]{WangYBCOthickness}
\bibinfo{author}{\bibfnamefont{T.}~\bibnamefont{Wang}} \bibnamefont{et~al.},
  \bibinfo{journal}{Phys. Rev. B} \textbf{\bibinfo{volume}{43}},
  \bibinfo{pages}{8623} (\bibinfo{year}{1991}).

\bibitem[{\citenamefont{Salluzzo
  et~al.}(2007{\natexlab{a}})\citenamefont{Salluzzo, De~Luca, and
  Vaglio}}]{SalluzzoNBCO}
\bibinfo{author}{\bibfnamefont{M.}~\bibnamefont{Salluzzo}},
  \bibinfo{author}{\bibfnamefont{G.}~\bibnamefont{De~Luca}}, \bibnamefont{and}
  \bibinfo{author}{\bibfnamefont{R.}~\bibnamefont{Vaglio}},
  \bibinfo{journal}{Applied Superconductivity, IEEE Transactions on}
  \textbf{\bibinfo{volume}{17}}, \bibinfo{pages}{3569 }
  (\bibinfo{year}{2007}{\natexlab{a}}).

\bibitem[{\citenamefont{Orgiani et~al.}(2007)}]{OrgianiPRLhighTcRc}
\bibinfo{author}{\bibfnamefont{P.}~\bibnamefont{Orgiani}} \bibnamefont{et~al.},
  \bibinfo{journal}{Phys. Rev. Lett.} \textbf{\bibinfo{volume}{98}},
  \bibinfo{pages}{036401} (\bibinfo{year}{2007}).

\bibitem[{\citenamefont{Ando et~al.}(1996)}]{AndoLSCOHighfield}
\bibinfo{author}{\bibfnamefont{Y.}~\bibnamefont{Ando}} \bibnamefont{et~al.},
  \bibinfo{journal}{Journal of Low Temperature Physics}
  \textbf{\bibinfo{volume}{105}}, \bibinfo{pages}{867} (\bibinfo{year}{1996}).

\bibitem[{\citenamefont{Steiner et~al.}(2005)\citenamefont{Steiner, Boebinger,
  and Kapitulnik}}]{SteinerHighTcSIhighField}
\bibinfo{author}{\bibfnamefont{M.~A.} \bibnamefont{Steiner}},
  \bibinfo{author}{\bibfnamefont{G.}~\bibnamefont{Boebinger}},
  \bibnamefont{and}
  \bibinfo{author}{\bibfnamefont{A.}~\bibnamefont{Kapitulnik}},
  \bibinfo{journal}{Phys. Rev. Lett.} \textbf{\bibinfo{volume}{94}},
  \bibinfo{pages}{107008} (\bibinfo{year}{2005}).

\bibitem[{\citenamefont{Seidler et~al.}(1992)\citenamefont{Seidler, Rosenbaum,
  and Veal}}]{SeidlerSIinYBCO}
\bibinfo{author}{\bibfnamefont{G.~T.} \bibnamefont{Seidler}},
  \bibinfo{author}{\bibfnamefont{T.~F.} \bibnamefont{Rosenbaum}},
  \bibnamefont{and} \bibinfo{author}{\bibfnamefont{B.~W.} \bibnamefont{Veal}},
  \bibinfo{journal}{Phys. Rev. B} \textbf{\bibinfo{volume}{45}},
  \bibinfo{pages}{10162} (\bibinfo{year}{1992}).

\bibitem[{\citenamefont{Doiron-Leyraud et~al.}(2006)}]{NicolasPRLYBCOSI1}
\bibinfo{author}{\bibfnamefont{N.}~\bibnamefont{Doiron-Leyraud}}
  \bibnamefont{et~al.}, \bibinfo{journal}{Phys. Rev. Lett.}
  \textbf{\bibinfo{volume}{97}}, \bibinfo{pages}{207001}
  (\bibinfo{year}{2006}).

\bibitem[{\citenamefont{Ahn et~al.}(1999)}]{AhnScience1999GBCO}
\bibinfo{author}{\bibfnamefont{C.~H.} \bibnamefont{Ahn}} \bibnamefont{et~al.},
  \bibinfo{journal}{Science} \textbf{\bibinfo{volume}{284}},
  \bibinfo{pages}{1152} (\bibinfo{year}{1999}).

\bibitem[{\citenamefont{Mannhart}(1996)}]{MannhartSSTreview}
\bibinfo{author}{\bibfnamefont{J.}~\bibnamefont{Mannhart}},
  \bibinfo{journal}{Superconductor Science and Technology}
  \textbf{\bibinfo{volume}{9}}, \bibinfo{pages}{49} (\bibinfo{year}{1996}).

\bibitem[{\citenamefont{Tanda et~al.}(1992)\citenamefont{Tanda, Ohzeki, and
  Nakayama}}]{TandaSIinNCCO}
\bibinfo{author}{\bibfnamefont{S.}~\bibnamefont{Tanda}},
  \bibinfo{author}{\bibfnamefont{S.}~\bibnamefont{Ohzeki}}, \bibnamefont{and}
  \bibinfo{author}{\bibfnamefont{T.}~\bibnamefont{Nakayama}},
  \bibinfo{journal}{Phys. Rev. Lett.} \textbf{\bibinfo{volume}{69}},
  \bibinfo{pages}{530} (\bibinfo{year}{1992}).

\bibitem[{\citenamefont{Ahn et~al.}(2006)}]{AhnRMP}
\bibinfo{author}{\bibfnamefont{C.~H.} \bibnamefont{Ahn}} \bibnamefont{et~al.},
  \bibinfo{journal}{Rev. Mod. Phys.} \textbf{\bibinfo{volume}{78}},
  \bibinfo{pages}{1185} (\bibinfo{year}{2006}).

\bibitem[{\citenamefont{R\"ufenacht
  et~al.}(2006)}]{RufenachtSupercluidDensityinLSCO}
\bibinfo{author}{\bibfnamefont{A.}~\bibnamefont{R\"ufenacht}}
  \bibnamefont{et~al.}, \bibinfo{journal}{Phys. Rev. Lett.}
  \textbf{\bibinfo{volume}{96}}, \bibinfo{pages}{227002}
  (\bibinfo{year}{2006}).

\bibitem[{\citenamefont{Matthey et~al.}(2007)\citenamefont{Matthey, Reyren,
  Triscone, and Schneider}}]{MattheyNBCOonSTO}
\bibinfo{author}{\bibfnamefont{D.}~\bibnamefont{Matthey}},
  \bibinfo{author}{\bibfnamefont{N.}~\bibnamefont{Reyren}},
  \bibinfo{author}{\bibfnamefont{J.-M.} \bibnamefont{Triscone}},
  \bibnamefont{and}
  \bibinfo{author}{\bibfnamefont{T.}~\bibnamefont{Schneider}},
  \bibinfo{journal}{Phys. Rev. Lett.} \textbf{\bibinfo{volume}{98}},
  \bibinfo{pages}{057002} (\bibinfo{year}{2007}).

\bibitem[{\citenamefont{Salluzzo et~al.}(2009)}]{SalluzzoSST}
\bibinfo{author}{\bibfnamefont{M.}~\bibnamefont{Salluzzo}}
  \bibnamefont{et~al.}, \bibinfo{journal}{Superconductor Science and
  Technology} \textbf{\bibinfo{volume}{22}}, \bibinfo{pages}{034010}
  (\bibinfo{year}{2009}).

\bibitem[{\citenamefont{Salluzzo et~al.}(2008{\natexlab{a}})}]{SalluzzoPRB2008}
\bibinfo{author}{\bibfnamefont{M.}~\bibnamefont{Salluzzo}}
  \bibnamefont{et~al.}, \bibinfo{journal}{Phys. Rev. B}
  \textbf{\bibinfo{volume}{78}}, \bibinfo{pages}{054524}
  (\bibinfo{year}{2008}{\natexlab{a}}).

\bibitem[{\citenamefont{Salluzzo
  et~al.}(2008{\natexlab{b}})}]{SalluzzoPRLNBCOCuOxchain}
\bibinfo{author}{\bibfnamefont{M.}~\bibnamefont{Salluzzo}}
  \bibnamefont{et~al.}, \bibinfo{journal}{Phys. Rev. Lett.}
  \textbf{\bibinfo{volume}{100}}, \bibinfo{pages}{056810}
  (\bibinfo{year}{2008}{\natexlab{b}}).

\bibitem[{\citenamefont{Salluzzo
  et~al.}(2007{\natexlab{b}})}]{SalluzzoPRB2007SI}
\bibinfo{author}{\bibfnamefont{M.}~\bibnamefont{Salluzzo}}
  \bibnamefont{et~al.}, \bibinfo{journal}{Phys. Rev. B}
  \textbf{\bibinfo{volume}{75}}, \bibinfo{pages}{054519}
  (\bibinfo{year}{2007}{\natexlab{b}}).

\bibitem[{\citenamefont{Ye et~al.}(2010)}]{YeNature}
\bibinfo{author}{\bibfnamefont{J.~T.} \bibnamefont{Ye}} \bibnamefont{et~al.},
  \bibinfo{journal}{Nat Mater} \textbf{\bibinfo{volume}{9}},
  \bibinfo{pages}{125} (\bibinfo{year}{2010}).

\bibitem[{\citenamefont{Lee et~al.}(2011)}]{YeonbaeSTO}
\bibinfo{author}{\bibfnamefont{Y.}~\bibnamefont{Lee}} \bibnamefont{et~al.},
  \bibinfo{journal}{Phys. Rev. Lett.} \textbf{\bibinfo{volume}{106}},
  \bibinfo{pages}{136809} (\bibinfo{year}{2011}).

\bibitem[{\citenamefont{Shimotani et~al.}(2007)}]{ShimotaniZnO}
\bibinfo{author}{\bibfnamefont{H.}~\bibnamefont{Shimotani}}
  \bibnamefont{et~al.}, \bibinfo{journal}{Applied Physics Letters}
  \textbf{\bibinfo{volume}{91}}, \bibinfo{eid}{082106} (\bibinfo{year}{2007}).

\bibitem[{\citenamefont{Dhoot et~al.}(2010)}]{DhootYBCO}
\bibinfo{author}{\bibfnamefont{A.~S.} \bibnamefont{Dhoot}}
  \bibnamefont{et~al.}, \bibinfo{journal}{Advanced Materials}
  \textbf{\bibinfo{volume}{22}}, \bibinfo{pages}{2529} (\bibinfo{year}{2010}).

\bibitem[{\citenamefont{Bollinger et~al.}(2010)}]{BollingerLSCO}
\bibinfo{author}{\bibfnamefont{A.~T.} \bibnamefont{Bollinger}}
  \bibnamefont{et~al.}, \bibinfo{journal}{accepted by Nature}
  (\bibinfo{year}{2010}).

\bibitem[{\citenamefont{Varela et~al.}(1999)}]{JacoboYBCOPRL1999}
\bibinfo{author}{\bibfnamefont{M.}~\bibnamefont{Varela}} \bibnamefont{et~al.},
  \bibinfo{journal}{Phys. Rev. Lett.} \textbf{\bibinfo{volume}{83}},
  \bibinfo{pages}{3936} (\bibinfo{year}{1999}).

\bibitem[{\citenamefont{Varela et~al.}(2001)}]{JacoboYBCOPRL2001}
\bibinfo{author}{\bibfnamefont{M.}~\bibnamefont{Varela}} \bibnamefont{et~al.},
  \bibinfo{journal}{Phys. Rev. Lett.} \textbf{\bibinfo{volume}{86}},
  \bibinfo{pages}{5156} (\bibinfo{year}{2001}).

\bibitem[{\citenamefont{Jaeger et~al.}(1989)\citenamefont{Jaeger, Haviland,
  Orr, and Goldman}}]{Allen2DGranular}
\bibinfo{author}{\bibfnamefont{H.~M.} \bibnamefont{Jaeger}},
  \bibinfo{author}{\bibfnamefont{D.~B.} \bibnamefont{Haviland}},
  \bibinfo{author}{\bibfnamefont{B.~G.} \bibnamefont{Orr}}, \bibnamefont{and}
  \bibinfo{author}{\bibfnamefont{A.~M.} \bibnamefont{Goldman}},
  \bibinfo{journal}{Phys. Rev. B} \textbf{\bibinfo{volume}{40}},
  \bibinfo{pages}{182} (\bibinfo{year}{1989}).

\bibitem[{\citenamefont{Spivak et~al.}(2008)\citenamefont{Spivak, Oreto, and
  Kivelson}}]{SpivakTheoryPRB}
\bibinfo{author}{\bibfnamefont{B.}~\bibnamefont{Spivak}},
  \bibinfo{author}{\bibfnamefont{P.}~\bibnamefont{Oreto}}, \bibnamefont{and}
  \bibinfo{author}{\bibfnamefont{S.~A.} \bibnamefont{Kivelson}},
  \bibinfo{journal}{Phys. Rev. B} \textbf{\bibinfo{volume}{77}},
  \bibinfo{pages}{214523} (\bibinfo{year}{2008}).

\bibitem[{\citenamefont{Chen et~al.}(2009)\citenamefont{Chen, Andersen, and
  Hirschfeld}}]{ChenRTupturnPRB2009}
\bibinfo{author}{\bibfnamefont{W.}~\bibnamefont{Chen}},
  \bibinfo{author}{\bibfnamefont{B.~M.} \bibnamefont{Andersen}},
  \bibnamefont{and} \bibinfo{author}{\bibfnamefont{P.~J.}
  \bibnamefont{Hirschfeld}}, \bibinfo{journal}{Phys. Rev. B}
  \textbf{\bibinfo{volume}{80}}, \bibinfo{pages}{134518}
  (\bibinfo{year}{2009}).

\bibitem[{\citenamefont{Das and Doniach}(1999)}]{DasDoniachPRBBosemetal}
\bibinfo{author}{\bibfnamefont{D.}~\bibnamefont{Das}} \bibnamefont{and}
  \bibinfo{author}{\bibfnamefont{S.}~\bibnamefont{Doniach}},
  \bibinfo{journal}{Phys. Rev. B} \textbf{\bibinfo{volume}{60}},
  \bibinfo{pages}{1261} (\bibinfo{year}{1999}).

\bibitem[{\citenamefont{Doiron-Leyraud et~al.}(2007)}]{YBCOphase}
\bibinfo{author}{\bibfnamefont{N.}~\bibnamefont{Doiron-Leyraud}}
  \bibnamefont{et~al.}, \bibinfo{journal}{Nature}
  \textbf{\bibinfo{volume}{447}}, \bibinfo{pages}{565} (\bibinfo{year}{2007}).

\bibitem[{\citenamefont{Herbut}(2001)}]{Herbutquantumcriticalpoint}
\bibinfo{author}{\bibfnamefont{I.~F.} \bibnamefont{Herbut}},
  \bibinfo{journal}{Phys. Rev. Lett.} \textbf{\bibinfo{volume}{87}},
  \bibinfo{pages}{137004} (\bibinfo{year}{2001}).

\bibitem[{\citenamefont{R\"uhl\"ander and Soukoulis}(2001)}]{MarcMItransition}
\bibinfo{author}{\bibfnamefont{M.}~\bibnamefont{R\"uhl\"ander}}
  \bibnamefont{and} \bibinfo{author}{\bibfnamefont{C.~M.}
  \bibnamefont{Soukoulis}}, \bibinfo{journal}{Phys. Rev. B}
  \textbf{\bibinfo{volume}{63}}, \bibinfo{pages}{085103}
  (\bibinfo{year}{2001}).

\bibitem[{\citenamefont{Lee et~al.}(1993)\citenamefont{Lee, Wang, and
  Kivelson}}]{DunghaiLeeQuantumPercolation}
\bibinfo{author}{\bibfnamefont{D.-H.} \bibnamefont{Lee}},
  \bibinfo{author}{\bibfnamefont{Z.}~\bibnamefont{Wang}}, \bibnamefont{and}
  \bibinfo{author}{\bibfnamefont{S.}~\bibnamefont{Kivelson}},
  \bibinfo{journal}{Phys. Rev. Lett.} \textbf{\bibinfo{volume}{70}},
  \bibinfo{pages}{4130} (\bibinfo{year}{1993}).

\end{thebibliography}
\end{document}